\documentclass[conference]{IEEEtran}

\IEEEoverridecommandlockouts
\usepackage[ruled,vlined,linesnumbered]{algorithm2e}
\usepackage{booktabs}
\usepackage{adjustbox}
\usepackage{url}
\usepackage{multirow}
\usepackage{amsmath, amssymb}
\usepackage{hyperref}
\usepackage{cleveref}
\usepackage{multicol}
\title{Empirical Analysis and Detection of Hallucinations in LLM-Generated Bug Report Summaries} 
\author{

\IEEEauthorblockN{1\textsuperscript{st} Hinduja Nirujan}
\IEEEauthorblockA{\textit{Electrical and Software Engineering} \\
\textit{University of Calgary}\\
Calgary, Alberta, Canada \\
hinduja.nirujan@ucalgary.ca}
\and
\IEEEauthorblockN{2\textsuperscript{nd} Shreyas Patil}
\IEEEauthorblockA{\textit{Electrical and Software Engineering} \\
\textit{University of Calgary}\\
Calgary, Alberta, Canada \\
shreyas.patil@ucalgary.ca}
\and
\IEEEauthorblockN{3\textsuperscript{rd} Abdallah Ayoub}
\IEEEauthorblockA{\textit{Electrical and Software Engineering} \\
\textit{University of Calgary}\\
Calgary, Alberta, Canada \\
abdallah.ayoub@ucalgary.ca}
\and
\IEEEauthorblockN{4\textsuperscript{th} Ahmad Abdel Latif}
\IEEEauthorblockA{\textit{Electrical and Software Engineering} \\
\textit{University of Calgary}\\
Calgary, Alberta, Canada \\
ahmad.abdellatif@ucalgary.ca}
\and
\IEEEauthorblockN{5\textsuperscript{th} Gouri Ginde}
\IEEEauthorblockA{\textit{Electrical and Software Engineering} \\
\textit{University of Calgary}\\
Calgary, Alberta, Canada \\
gouri.ginde@ucalgary.ca}
}

\begin{document}
\maketitle
\begin{abstract}
\textit{Background:} Large Language Models (LLMs) are increasingly used to generate summaries of software bug reports, including sections such as Steps-to-Reproduce (S2R), Actual Behavior (AB), and Expected Behavior (EB). However, these models frequently produce hallucinations that can be convincing but are unsupported by the source bug report. This can mislead developers and reduce trust in automated software maintenance tools. Existing hallucination detection approaches typically evaluate outputs at the full-response level and do not consider the structural properties of technical documents. An initial exploratory study on 80 structured bug report summaries revealed that approximately 47.9\% contained missing information, while 12.3\% included fabricated content, highlighting the need for systematic hallucination analysis in bug report summarization. 
\textit{Aims:} In this work, we empirically investigate hallucinations in LLM-generated bug report summaries from a section-aware perspective. 
\textit{Method:} Using the BugsRepo dataset (multifaceted dataset
derived from Mozilla OSS projects), we introduce controlled synthetic hallucination injection to construct a benchmark for training and evaluation. We propose a section-aware hallucination detection approach that jointly predicts whether a summary contains hallucinated content, identifies the affected bug report sections, and classifies hallucination types. 
\textit{Results:} Experimental results across multiple pretrained language models show that the proposed approach achieves strong performance across all tasks, with the best model obtaining 0.89 report-level Macro-F1, 0.83 section-level Macro-F1, and 0.84 hallucination-type Macro-F1. We further analyze common hallucination patterns and model failure modes to better understand the limitations of current LLM-generated bug report summaries. 
\textit{Conclusions:} The findings highlight the importance of section-aware hallucination analysis for improving the reliability of LLM-assisted bug report summarization in software maintenance workflows. 
\end{abstract}

\section{Introduction}
\label{sec:Intro}
Software maintenance is a major expense in the software development process, costing billions of dollars every year \cite{jorgensen_shepperd_2007}. For example, the Consortium for Information \& Software Quality estimated that poor software quality cost the U.S. at least \$2.41 trillion in 2022, with accumulated technical debt \cite{cisq_2022}. The major hurdle in the software development industry is finding and fixing software issues, including bugs, which can affect functionality and security vulnerabilities that can put the system at risk. This effort is substantial as prior work reports that developers spend, on average, about 50\% of their time finding and fixing bugs \cite{HAMILL20171}.   

The core of bug resolution relies on the bug reports, but their effectiveness relies on the details present in the bug report. Well-structured reports are clearly able to articulate actual behavior(AB), expected behavior(EB), and steps to reproduce (S2R). This structuring minimizes ambiguity and enables developers to resolve issues without much discussion and clarification \cite{wang2015evaluating}. With recent advancements in Natural Language Processing (NLP) and and Large Language Models (LLMs) showed strong capabilities of in various Software Engineering (SE) tasks such as bug report generation and summarization \cite{bug_report_llm_2025}. The models have demonstrated significant potential in assisting software development and maintenance activities. However, despite their promising performance, LLMs are prone to generating plausible-sounding yet factually incorrect information, commonly referred to as \textit{hallucinations}. This phenomenon poses a significant challenge to the reliability and trustworthiness of LLM outputs, particularly in critical SE tasks where accuracy and faithfulness to the source information are essential. Therefore, detecting and mitigating hallucinations in LLM-generated output (e.g., bug reports) is crucial for ensuring their practical adoption in real-world SE workflows.

\begin{figure}[t]
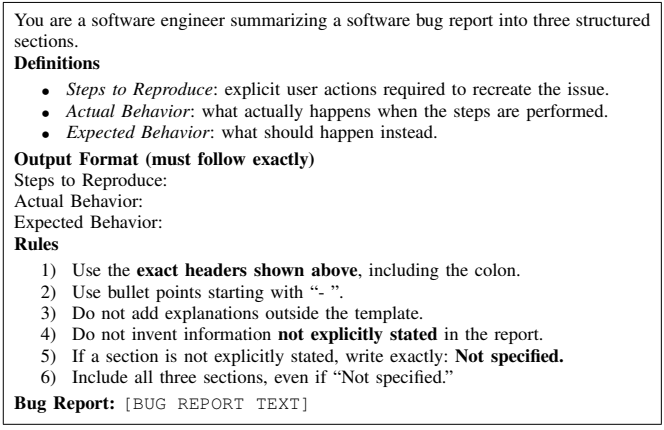

\centering
\setlength{\fboxsep}{4pt}
\setlength{\fboxrule}{0.3pt}
\fbox{%
\begin{minipage}{0.95\columnwidth}
\scriptsize
You are a software engineer summarizing a software bug report into three structured sections. \\
\textbf{Definitions}
\begin{itemize}
    \item \textit{Steps to Reproduce}: explicit user actions required to recreate the issue.
    \item \textit{Actual Behavior}: what actually happens when the steps are performed.
    \item \textit{Expected Behavior}: what should happen instead.
\end{itemize}
\textbf{Output Format (must follow exactly)} \\
\noindent Steps to Reproduce:\\
Actual Behavior:\\
Expected Behavior: \\
\textbf{Rules}
\begin{enumerate}
    \item Use the \textbf{exact headers shown above}, including the colon.
    \item Use bullet points starting with ``- ''.
    \item Do not add explanations outside the template.
    \item Do not invent information \textbf{not explicitly stated} in the report.
    \item If a section is not explicitly stated, write exactly: \textbf{Not specified.}
    \item Include all three sections, even if ``Not specified.''
\end{enumerate}
\textbf{Bug Report:}
\noindent \texttt{[BUG REPORT TEXT]}
\end{minipage}%
}
\caption{Prompt template used to generate structured bug report summaries.}
\label{fig:summary-generation-prompt}
\vspace{-7mm}
\end{figure}

Existing hallucination detection methods can be broadly classified into: self-consistency \cite{SelfCheckGPT_2025}, knowledge-graph two-stage detection \cite{kale2025liemeknowledgegraphs}, fine-tuned classifiers, and LLM-as-a-Judge \cite{gu2024survey}. While these techniques have advanced the field, there has been little research specifically examining hallucinations in the SE domain. Moreover, these methods treat hallucinations as a whole and monolithically, considering them as a single class of errors. This perspective fails to capture the diversity of factual errors and overlooks how hallucination patterns can vary across different semantic sections of a bug report. As a result, these methods do not account for the distinct roles and contextual behaviors of sections. Understanding hallucination patterns is crucial, as different types of hallucinations may require different detection and mitigation strategies depending on the semantic context and functional role of each section in the bug report. \\
\textbf{Motivation. }To better understand the behavior of hallucinations in LLM-generated bug report summaries, we conducted an exploratory analysis on a subset of the BugsRepo dataset \cite{BugsRepo_2025}, a multifaceted curated dataset derived from Mozilla projects of bug reports organized into semantic fields such as S2R, AB, and EB. Specifically, we randomly selected 80 bug reports from the BugsRepo unstructured dataset, based on their structural characteristics and length. Next, we prompted Llama 3.2 using a prompt that contains explicit definitions and rules as shown in Fig.\ref{fig:summary-generation-prompt}, with the temperature set to 0.0 to ensure deterministic outputs. The summaries were generated under two compression settings according to the report length distribution, where low compression allowed up to 180 tokens and high compression allowed 90 tokens. Each summary was then independently and anonymously annotated by three authors using three labels: supported, incomplete, and fabricated. The agreement between the annotators was measured using Cohen’s Kappa \cite{cohens_kappa}, giving a high agreement (Kappa = 0.85). After resolving disagreements through discussion among the annotators, the final labels were established. 

\begin{figure}[t]
\centering
\includegraphics[width=0.9\columnwidth]{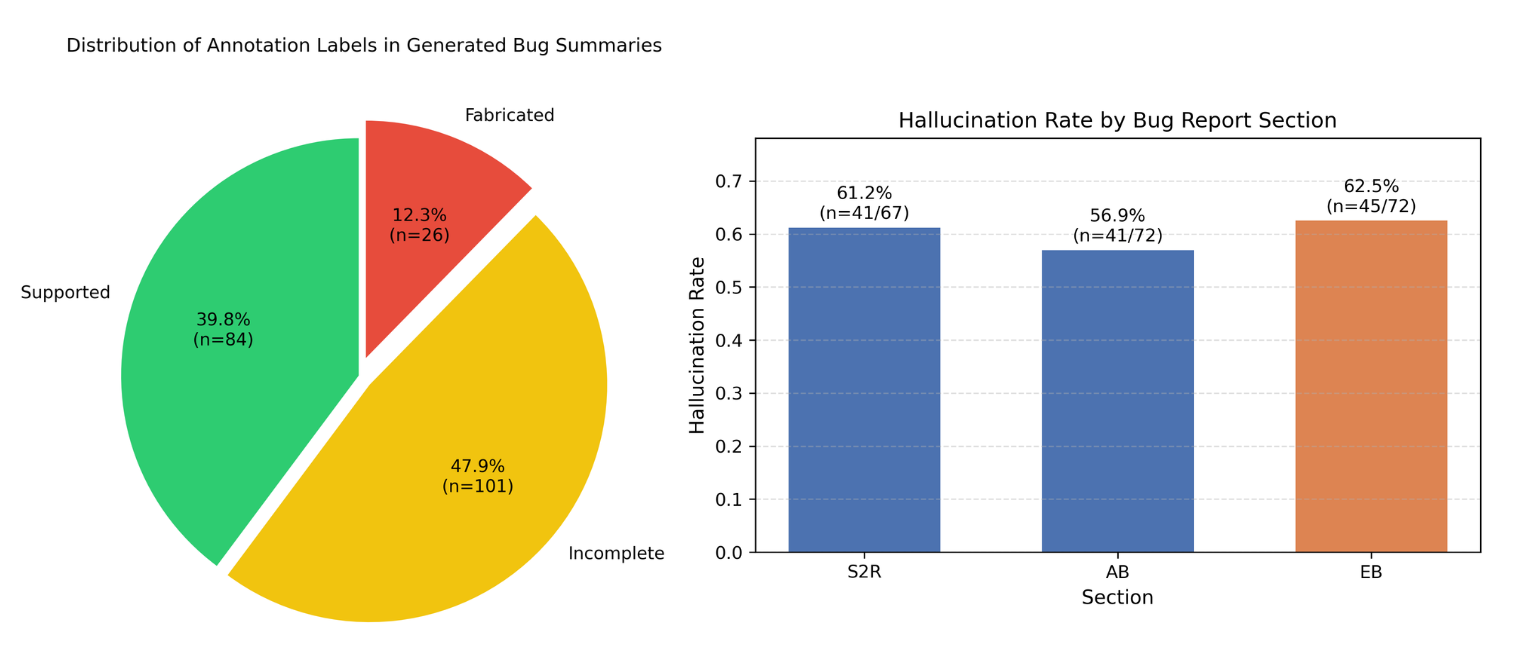}
\caption{Results of the exploratory analysis of hallucinations in LLM-generated bug report summaries.}
\label{fig:eda-results}
\vspace{-6mm}
\end{figure}

Figure \ref{fig:eda-results} presents the results of this analysis. From the figure, we observe that hallucinations are prevalent in LLM-generated bug report summaries, with over 60\% of summary segments containing either incomplete or fabricated information. Further, hallucination rates vary across semantic sections of bug reports, suggesting that hallucination behavior is not uniform and may depend on the contextual role of each section.

To address this gap, we propose a section-aware hallucination detection approach for structured bug report summaries. Using BugsRepo \cite{BugsRepo_2025} subset of structured reports as the source corpus, we constructed a synthetic hallucination dataset by injecting controlled inconsistencies into specific semantic sections, then fine-tuned language models to detect hallucinated sections and classify hallucination types.  More specifically, our study answers the following three research questions (RQs) that progressively address the key components required for this approach:
\begin{itemize}
    \item \textbf{RQ1:} \textit{How does hallucination sensitivity vary across different sections of structured bug reports during natural language generation?}
    Before defining our methodology, we analyze how transformer models distribute attention across different sections. These insights provide a foundation for our synthetic dataset distribution and help validate the empirical behaviors observed during zero-shot prompting, increasing confidence in our results.
    \item \textbf{RQ2:} \textit{How can structured bug reports be reliably converted into high-quality unstructured reports for training hallucination detection models?}
    Prior work typically evaluates structured-to-unstructured report conversions using semantic similarity metrics like cosine similarity~\cite{BugsRepo_2025}. However, our experiments show that these metrics fail to capture the core technical essence of structured reports, highlighting the need for a more reliable verification pipeline.
    \item \textbf{RQ3:} \textit{How effectively can a multi-task learning model detect hallucinations, identify affected sections, and classify hallucination types in bug report summaries?}
    An effective hallucination guardrail must first detect if an error exists before identifying the affected sections and hallucination types. This hierarchical dependency requires a multi-task learning approach, enabling the model to learn that section localization and type classification are conditional upon primary error detection.
\end{itemize}

    Through these RQs, our goal is to empirically study hallucinations in LLM-generated bug report summaries, analyze section-level vulnerability, construct a filtered benchmark of structured-to-unstructured conversions with controlled hallucination injection, and evaluate a multitask detector for report-, section-, and type-level hallucination detection.
    Our specific contributions are as follows:
\begin{enumerate}
    \item We introduce a novel section-aware approach for analyzing hallucinations in LLM-generated bug report summaries by explicitly modeling the semantic roles of S2R, AB, and EB for jointly detecting hallucinated summaries, identifying affected sections, and classifying hallucination types.
    \item We develop a synthetic dataset construction pipeline that converts structured bug reports into unstructured generated bug reports and creates hallucination-injected structured summaries to enable systematic section-level hallucination experiments. We make this benchmarking dataset publicly available for researchers to use and experiment \footnote{https://anonymous.4open.science/r/Anon\_Bug\_Report\_Summarization-1477}.  
    \item We provide empirical insights into hallucination behavior across bug report sections, highlighting patterns that can inform the development of more reliable LLM-assisted bug reporting tools.
\end{enumerate}
\textbf{Paper Organization.} Section~\ref{sec:RelatedWork} reviews the related work. Section~\ref{sec:Methodology} describes the dataset construction process, model setup, and experimental design. The Section~\ref{sec:Results} explains the analysis and answers to each RQ. Section~\ref{sec:Discussion} presents the discussion and outlines directions for future work. Section~\ref{sec:Threats} discusses the threats to validity, and Section~\ref{sec:Conclusion} concludes the paper.

\section{Related Work}
\label{sec:RelatedWork}
In this section, we will present work related to bug report summarization, and hallucination detection in SE artifacts. \\
\textbf{Bug Report Summarization.}
Automatic bug report summarization aims to condense lengthy and noisy issue discussions into concise representations that help developers understand and resolve bugs efficiently. Early work emphasized content-aware and noise-aware unsupervised summarization methods for bug reports. Mani et al. proposed AUSUM, an unsupervised approach that reduce noise sentences from bug reports by considering sentence structure and topic relevance\cite{AUSUM_2012}. Lotufo et al. modeled the 'hurried' bug report reading process and ranked sentences according to discussed topics, evaluation and report title signals \cite{HurriedBug}. Jiang et al. proposed PRST, a PageRank-based summarization technique that considers duplicate bug reports, showing that duplicate discussions can provide useful evidence for identifying important report content \cite{PRST_2017}. More recent approaches adopt pre-trained representations and contrastive objectives to improve bug report summarization. RepresentThemAll learns a universal bug report representation for several downstream tasks, including summarization\cite{representThemAll_2023}.KSCLP adds knowledge-specific pre-training before Seq2Seq summary generation \cite{knowledgeContrastiveSum_2024}, and SumLLaMA adapts LLMs using component-level contrastive pre-training and LoRA-based fine-tuning \cite{sumllama_2024}. \\
While these approaches improve overall summary generation, recent studies have started to focus on structured bug report summaries. Mukhtar et al. proposed ClaSum, a multidocument summarization approach that combines classification and summarization to cover key bug attributes such as bug description, reproduction steps, environment, and solutions \cite{multi_document_2024}. Acharya and Ginde studied whether instruction-tuned LLMs can transform casual, unstructured bug reports into structured reports containing fields such as S2R, EB, and AB \cite{bug_report_llm_2025}. BRMDS further formulates bug report summarization as a multi-dimensional generation task, producing summaries across environment, actual behavior, expected behavior, bug category, and solution suggestions using PEFT-tuned LLMs \cite{brmds_2025}. However, the existing work primarily evaluates summary informativeness, semantic similarity, and overall generation quality. They do not explicitly assess whether each generated section is factually grounded in the source bug report. This motivates our study of hallucination detection in structured bug report summaries, especially at the section level.\\ 
\textbf{Hallucination Detection in SE Artifacts.}
Hallucinations in LLMs refer to outputs that are factually incorrect, unsupported by context, or logically inconsistent. This problem has been widely studied in open-domain NLP tasks such as question answering, dialogue generation, and text summarization \cite{summary-faithfulness, hallucination_survey, hallucination_review_2026}. Recent work includes self-consistency, model-based evaluation, and taxonomy-driven analyses. SelfCheckGPT uses self-consistency by sampling multiple responses to the same prompt and detecting hallucinations from factual inconsistencies across those samples \cite{SelfCheckGPT_2025}. SelfCheckEval extends this with black-box, semantic and contextual consistency checks \cite{SelfCheckEval_2025}, while taxonomy-driven studies categorize different hallucination types \cite{Hallucinot_2025}. However, hallucination detection in SE requires additional attention because LLM outputs may become artifacts that developers inspect, reuse, or execute. \\
Recent SE research shows that hallucinations occur across both executable and natural-language software artifacts. In code generation, hallucinations may appear as syntactically plausible but incorrect code, fabricated APIs, nonexistent dependencies, or logic that violates the intended functionality. Prior work characterizes code hallucinations through factuality, faithfulness, and compatibility, covering API errors, requirement deviations, and dependency or environment assumptions \cite{code-hallucination-LR, SE-LLM-survey, functional_correctness, code-gen-hallucination}. Package hallucination studies show that LLMs can generate nonexistent package names, creating software supply-chain risks \cite{package-hallucination, hallucination-inspector}. Recent research also studies hallucinations in code-to-natural-language artifacts. ETF \cite{etf-tracing} detects hallucinations in code summaries by tracing generated entities back to source code. Liu et al. show that generated code reviews and commit messages can contain input inconsistencies, logic inconsistencies, and intent deviations \cite{code-change-hallucinations}. Similarly, HalluJudge detects whether generated review comments are grounded in the corresponding code diff without the reference comments \cite{HalluJudge}.  \\
These studies show that SE hallucination detection requires a grounding generated text in technical evidence such as source code, code diffs, APIs, etc. Handling bug reports  can be a challenge as they combine natural language with structured technical evidence. The CHIME approach \cite{chatGPT_analysis_2025} studies inaccuracies in ChatGPT responses to bug-report questions and shows that LLMs struggle with stack traces, metadata, and context distributed across issue-report text. This is relevant to our setting because bug report hallucination detection must account for technical artifacts embedded in natural-language reports. These studies highlight the unique challenges of the SE context, where factual grounding depends on structured technical elements rather than purely natural language. Unlike prior work that focuses on whole-response hallucination detection or code-oriented artifacts, our work specifically examines whether semantic sections such as S2R, AB, and EB are factually grounded in the source bug report. \\
Our work investigates section-level hallucination in structured bug report summaries, where each semantic section must be grounded in the corresponding source report. This approach enables a more fine-grained evaluation of hallucination detection methods for SE artifacts.
\section{Methodology}
\label{sec:Methodology}
\begin{figure*}[t]
    \centering
    \includegraphics[scale=.35 ]{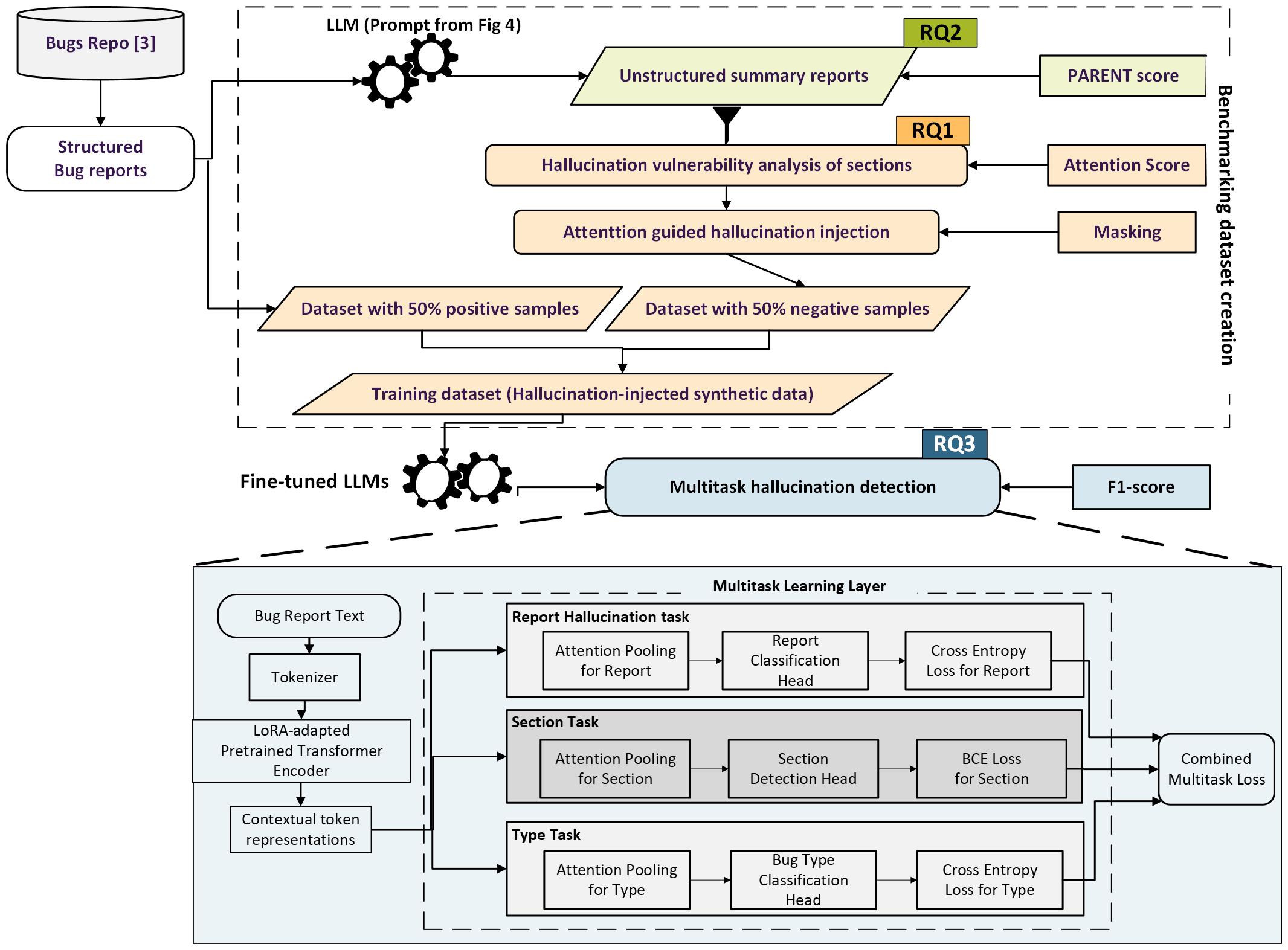}
    \caption{Our proposed section-aware hallucination detection approach and study design}
    \label{fig:system-diagram}
\end{figure*}

The overall methodology of the proposed section-aware hallucination detection approach is shown in Fig.\ref{fig:system-diagram}.

\textbf{Dataset.} 
BugsRepo \cite{BugsRepo_2025} is a curated dataset derived from Mozilla Bugzilla projects, containing bug reports along with metadata, comments, and contributor information. Relevant to our work, BugsRepo includes a structured subset of 10,351 well-formed reports with S2R, AB, and EB fields. Since our study focuses on section-aware hallucination in structured bug report summaries, we use this structured subset. After preprocessing and removing unusable records, our experiments use 10,304 structured bug reports created between January 2021 and November 2024 for section attention analysis, structured-to-unstructured report conversion, and synthetic hallucination injection components.

\subsection{Section Attention Analysis} 
\textbf{Section Extractor.}
In the BugsRepo dataset, bug reports are provided in a semi-structured format containing three primary components: S2R, AB, and EB. Despite this structural organization, the reports frequently include extraneous content such as human commentary, stack traces, and references to GitHub pull requests. Incorporating such artifacts into section-level analysis can introduce noise and bias the distribution of attention, as irrelevant tokens may dilute or overshadow semantically meaningful information within each section.
To address this challenge, we propose a deterministic, rule-based extraction approach that isolates the core semantic components of bug reports. The study leverages lexical pattern matching and structural heuristics to accurately identify and segment relevant sections while filtering out non-informative content. It is specifically designed to be robust against formatting inconsistencies and the noisy, heterogeneous nature of real-world bug reports, thereby preserving the integrity of section-level representations for downstream analysis.\\
\textbf{Attention Analysis.} After isolating individual \texttt{S2R}, \texttt{AB}, and \texttt{EB} sections, our objective is to empirically evaluate how transformer architectures distribute contextual importance across these distinct semantic regions. To achieve this, we analyze intra-section and inter-section attention dynamics, leveraging multi-head self-attention weights to quantify information routing and cross-field dependencies within the bug reports. Our entire process for analyzing and aggregating multi-model attention distributions is outlined in \cref{alg:attention}.
\newline To ensure a broad coverage in terms of different text-processing models, we consider multiple prominent pretrained transformers, such as BERT, RoBERTa, DistilBERT, and even domain-adapted models, like SciBERT, and DeBERTa, which have been shown to be successful in analyzing contextual relationships within software artifacts~\cite{transformers, BugsRepo_2025}. A key feature of the mentioned models is their multi-head self-attention architecture, which allows us to measure token-to-token interactions. For each bug report, we create a single sequence of tokens that explicitly encodes the extracted section borders:
\begin{equation}
\begin{aligned}
T &=
[\texttt{CLS}] \; [\texttt{S2R}] \; T_{s2r} \;
[\texttt{AR}] \; T_{ar} \;
[\texttt{ER}] \; T_{er} \; [\texttt{SEP}], \\
\mathcal{I} &= \{I_{s2r}, I_{ar}, I_{er}\}.
\end{aligned}
\end{equation}
where $T_{s2r}$, $T_{ar}$, and $T_{er}$ denote the tokenized representations of the S2R, AB, and EB sections, respectively. The index sets in $\mathcal{I}$ enable section-level analysis by isolating tokens belonging to each section.

These index mappings allow precise isolation of tokens belonging to each section. Given an input sequence of length $N$, transformer models compute attention using multi-head self-attention across $L$ layers and $H$ attention heads. For each layer $l$, the attention tensor is defined as:
\begin{equation}
\begin{aligned}
A^{(l)} &\in \mathbb{R}^{H \times N \times N}, \\
A &= \frac{1}{L} \sum_{l=1}^{L}
\left(
    \frac{1}{H} \sum_{h=1}^{H} A^{(l)}_h
\right).
\end{aligned}
\end{equation}
where $H$ is the number of attention heads and $N$ is the number of tokens. Each head captures token-to-token interactions, where $A^{(l)}_h[i,j]$ represents the attention weight from token $i$ to token $j$ for head $h$ in layer $l$. The resulting matrix $A \in \mathbb{R}^{N \times N}$ encodes the average attention distribution across all token pairs. To quantify intra-section interactions, we compute the average attention between tokens within the same section. For a section $S \in \{s2r, ar, er\}$, the attention score is defined as:
\begin{equation}
\begin{aligned}
\mathrm{Attn}(S) &=
\frac{1}{|I_S|^2}
\sum_{i \in I_S}
\sum_{j \in I_S}
A[i,j], \\
\mathrm{Attn}(S \rightarrow T) &=
\frac{1}{|I_S| \cdot |I_T|}
\sum_{i \in I_S}
\sum_{j \in I_T}
A[i,j].
\end{aligned}
\end{equation}
This metric captures the internal coherence of each section. To analyze inter-section dependencies, we compute directional attention scores between different sections. For two distinct sections $S$ and $T$, the cross-section attention score is defined using ${Attn(S \rightarrow T)}$. This formulation captures the extent to which tokens in section $S$ attend to tokens in section $T$. The overall workflow for calculating these metrics across the base architectures is formalized in Algorithm~\ref{alg:attention}.

\begin{algorithm}[t]
\caption{Section-wise and Cross-section Attention Analysis}
\label{alg:attention}

\KwIn{Bug report dataset $D$ with sections S2R, AR, and ER}
\KwOut{Section-wise and cross-section attention scores}

$M \leftarrow \{\text{BERT}, \text{RoBERTa}, \text{SciBERT}, \text{DeBERTa}, \text{DistilBERT}\}$\;

\ForEach{bug report $d \in D$}{
    Extract and clean sections $s2r$, $ar$, and $er$\;
    Construct token sequence
    $T \leftarrow [\texttt{CLS}], [\texttt{S2R}], T_{s2r}, [\texttt{AR}], T_{ar}, [\texttt{ER}], T_{er}, [\texttt{SEP}]$\;
    Identify section index sets $I_{s2r}$, $I_{ar}$, and $I_{er}$\;

    \ForEach{model $m \in M$}{
        Obtain attention tensors $A^{(l)}$ for all layers\;
        Compute the averaged attention matrix
        $
        A \leftarrow \frac{1}{L} \sum_{l=1}^{L}
        \left(
            \frac{1}{H} \sum_{h=1}^{H} A^{(l)}_h
        \right)
        $\;

        \ForEach{section $S \in \{s2r, ar, er\}$}{
            Compute intra-section attention
            $
            \mathrm{Attn}(S) \leftarrow
            \frac{1}{|I_S|^2}
            \sum_{i \in I_S}
            \sum_{j \in I_S}
            A[i,j]
            $\;
        }

        \ForEach{ordered pair $(S,T)$ such that $S,T \in \{s2r, ar, er\}$ and $S \neq T$}{
            Compute cross-section attention
            $
            \mathrm{Attn}(S \rightarrow T) \leftarrow
            \frac{1}{|I_S|\,|I_T|}
            \sum_{i \in I_S}
            \sum_{j \in I_T}
            A[i,j]
            $\;
        }

        Store attention scores for model $m$\;
    }
}

\Return{Aggregated attention dataset}\;
\end{algorithm}
\subsection{Synthetic Dataset Construction} 
As there are no available datasets for hallucinations in section-aware bug report summaries, we constructed a benchmark dataset from the structured subset of the BugsRepo dataset.\\
\textbf{Unstructured Report Conversion.}
The structured reports are converted into unstructured reports using the Mistral-7B model using a refined prompt \ref{fig:structured-to-unstructured-prompt}. Since automatically generated reports may contain hallucinated or missing information, it is necessary to filter unreliable generations before constructing the training dataset. To evaluate this, we employ the PARENT (Precision and Recall of Entailed N-grams from the Table) metric \cite{parent}. PARENT evaluates table-to-text generation by measuring the agreement 
between the generated text $G$, the reference text $R$, and the source table $T$. 
The metric computes entailed precision and entailed recall over n-grams, rewarding 
generated tokens that either appear in the reference text or can be inferred from 
the source table. The overall PARENT score is defined as the harmonic mean of 
entailed precision ($E_p$) and entailed recall ($E_r$):
\begin{equation}
\mathrm{PARENT}(G,R,T) =
\frac{2 \cdot E_p \cdot E_r}{E_p + E_r}.
\end{equation}
However, a complete reference description for each structured bug report is not available in the dataset. It contains a summary field that provides a brief summary of the reports. Therefore, relying on reference-based recall may underestimate the correctness of generated content that faithfully reflects the structured report. To address this limitation, we introduce a simplified variant of the metric that prioritizes alignment with the source table. Specifically, we compute a modified score defined as:
\begin{equation}
\mathrm{PARENT}_{table} =
0.5 \cdot E_p + 0.5 \cdot E_r(T).
\label{eq:parent_table}
\end{equation}
where $E_p$ denotes the \emph{entailed precision} from the original PARENT metric. In contrast, $E_r(T)$ corresponds only to the recall from the source table, measuring how much of the structured report content is covered in the generated text, unlike the original PARENT formulation, which combines recall from both the reference text and the source table. The reasons and empirical evaluation of this modification are discussed further in Section~\ref{sub:RQ2}.
\begin{figure}[t]
\centering
\setlength{\fboxsep}{3pt}
\setlength{\fboxrule}{0.3pt}
\fbox{%
\begin{minipage}{0.97\columnwidth}
\scriptsize

\textbf{Prompt used for structured to unstructured report conversion}
Rewrite the following structured bug report as one coherent paragraph using the original wording as much as possible.
\textbf{Rules:}
\begin{itemize}
    \item Preserve the original meaning exactly.
    \item Do not invent, infer, summarize, or generalize.
    \item Do not drop important details.
    \item Keep all technical identifiers and quoted strings unchanged.
    \item Keep the distinction between expected behavior and actual behavior explicit.
    \item Use a single paragraph only.
    \item Make it readable, but stay as close as possible to the source wording.
\end{itemize}

\textbf{Structured bug report:}

\medskip

\noindent\texttt{[BUG REPORT TEXT]}

\end{minipage}%
}
\caption{Prompt template used to convert structured bug reports to unstructured reports.}
\label{fig:structured-to-unstructured-prompt}
\vspace{-3mm}
\end{figure}
Following the filtering procedure described in Section~\ref{sub:RQ2}, low-quality conversions were removed to ensure reliable structured-to-text alignment. After filtering, 7,684 high-quality unstructured bug reports were retained from the original 10,304 reports. From this filtered dataset, approximately 50\% of the samples (3,800 reports) were selected for synthetic hallucination injection, while the remaining reports were preserved as non-hallucinated instances. \\
\textbf{Synthetic Hallucination Injection}
 The annotated datasets for studying hallucinations and factual inconsistencies are limited. Creating this data manually is both time-consuming and prone to human error. Prior work has used synthetic perturbations to simulate hallucinated outputs and augment hallucination detection data \cite{summary-faithfulness, malto_2024}. Based on this, we design a controlled hallucination injection pipeline designed to systematically perturb structured bug report fields using an LLM. The objective is to generate semantically altered yet syntactically valid variants under strict constraints.
\begin{algorithm}[t]
\caption{Hallucination Injection Pipeline}
\label{alg:hallucination-injection}

\KwIn{Dataset $D$, maximum number of retries $K$}
\KwOut{Injected hallucination examples with success flags and attempt counts}

\ForEach{report $r \in D$}{
    \If{$r$ is marked for hallucination}{
        Select field $f$ and hallucination type $h$\;

        \For{$k \leftarrow 1$ \KwTo $K$}{
            Generate $f'_k \leftarrow \mathrm{LLM}(f, p_h)$\;
            Clean output $f'_k$\;

            \If{$\mathrm{Validate}(f, f'_k, h)$}{
                Accept $f' \leftarrow f'_k$\;
                \textbf{break}\;
            }
        }

        \If{no valid output is found}{
            Assign $f' \leftarrow f'_K$ and mark as failed\;
        }

        Store $f'$, success flag, and attempts\;
    }
}
\end{algorithm}

The hallucination injection process begins by iterating over each bug report in the dataset (Lines 1--2), where only those marked for hallucination are selected for further processing. For each such instance, the corresponding field and hallucination type are identified (Line 3). The pipeline then enters an iterative generation phase (Lines 4--8), where the LLM produces candidate outputs conditioned on the selected prompt. Each generated output is cleaned and validated against strict constraints specific to the hallucination type. The iteration terminates early if a valid transformation is obtained, ensuring efficiency while maintaining correctness. In cases where none of the generated outputs satisfy the validation criteria, the final attempt is retained and explicitly marked as a failure (Lines 9--11). Subsequently, the resulting output, along with metadata such as success status and number of attempts, is recorded (Line 12). To ensure robustness and fault tolerance, the pipeline periodically saves intermediate results, enabling recovery and seamless continuation in case of interruptions (Line 14).
\subsection{Section-Aware Hallucination Detection }
In the Section Aware detection, we evaluate multiple frontier LLMs; the previous approaches have shown that fine-tuned LLMs have higher performance \cite{bug_report_llm_2025} in terms of bug report summarization. From the synthetic data, we split the data for fine-tuning and testing the fine-tuned LLM. For each instance, we evaluate the LLM detection with metrics like accuracy, precision, F1-score, and recall.\\
\textbf{Multitask Hallucination Detector} 
Existing hallucination detection approaches \cite{SelfCheckGPT_2025, kale2025liemeknowledgegraphs, gu2024survey, Hallucinot_2025, SelfCheckEval_2025} typically frame the problem as a binary classification task that predicts whether a generated text contains hallucinated content. While this formulation is adequate for general text generation tasks, it is insufficient for structured bug reports. In bug reports, hallucinations may occur within specific sections and may be evident in different forms, such as additions, omissions, or reordering of structured information. Therefore, detecting the presence of hallucination alone does not provide sufficient insight into the location or type of the error.

To address this limitation, we propose a section-aware hallucination detection approach that simultaneously predicts (i) whether a report contains hallucinated content, (ii) which section of the bug report is affected, and (iii) the type of hallucination. The model learns three related tasks: a report-level hallucination classification task, a section-level localization task that identifies hallucinations in the S2R, AB, or EB sections, and a hallucination type classification task that distinguishes between \textit{addition}, \textit{removal}, and \textit{reordering} errors based on the results of the initial exploratory analysis explained in \ref{sec:Intro}.
As described in Figure \ref{fig:system-diagram}, the model is built upon a pretrained transformer encoder that generates contextual token representations for the input bug report. Task-specific representations are obtained using attention-based pooling layers, which aggregate contextual token representations relevant to each task. These pooled representations are then passed to three prediction heads: a binary classifier for report-level hallucination detection, a multi-label classifier for section-level localization, and a multi-class classifier for hallucination type prediction.
The model is trained using a multitask objective that combines the losses from the three tasks:
\begin{equation}
\mathcal{L}
=
\mathcal{L}_{report}
+ \lambda_s \mathcal{L}_{section}
+ \lambda_t \mathcal{L}_{type}.
\end{equation}
where $\lambda_s$ and $\lambda_t$ control the relative importance of the auxiliary tasks. To reduce computational cost, parameter-efficient fine-tuning (PEFT) is applied using Low-Rank Adaptation (LoRA), which updates small trainable adapter parameters in the transformer attention layers, together with the task-specific pooling layers and prediction heads. 

\section{Results}
\label{sec:Results}
After detailing a section-aware hallucination detection approach, we describe the methodology and present the results for each research question in this section.

\subsection{Answering RQ1: Hallucination sensitivity}
\label{sub:RQ1}
To analyze hallucination sensitivity across different sections of structured bug reports, we examine the average attention distribution across different sections of bug reports, namely S2R, AB, and EB, using multiple transformer models.
\begin{figure}[!htpb]
\centering
\includegraphics[width=0.9\columnwidth]{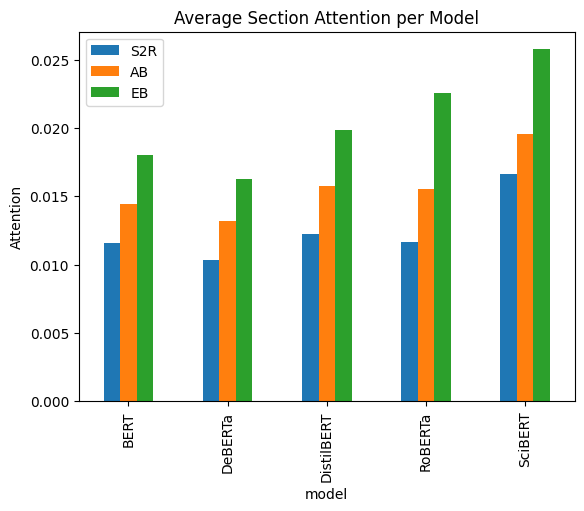}
\caption{Average section-wise attention across transformer models.}
\label{fig:attention}
\vspace{-3mm}
\end{figure}
As shown in Figure~\ref{fig:attention}, a consistent trend is observed across all models: EB receives the highest attention. AB receives moderate attention, and S2R receives the lowest attention.
Attention reflects the relative importance assigned to different sections during representation learning. The lower attention assigned to S2R indicates that models under-utilize procedural and sequential information, making it more susceptible to hallucination during generation. In contrast, EB sections, which are typically concise and declarative, receive higher attention and are less prone to hallucination.
\begin{table}[!htpb]
\centering
\caption{Distribution of injected hallucination samples by bug report section and hallucination type.}
\label{tab:hallucination-distribution}
\begin{tabular}{lrrr}
Section & Add & Remove & Reorder \\
\midrule
EB & 600 & 600 & -- \\
AB & 500 & 500 & -- \\
S2R & 584 & 583 & 433 \\
\end{tabular}
\end{table}
Guided by these observations, we construct a hallucination-sensitive dataset (half of the original dataset) by prioritizing sections with lower attention. The overall distribution is, S2R: 1600 samples, EB: 1200 samples, and AB: 1000 samples. Each section is further divided into hallucination types, as shown in the Table \ref{tab:hallucination-distribution}.
\begin{center}
\vspace{-4mm}
\setlength{\fboxsep}{6pt}
\setlength{\fboxrule}{0.4pt}
\fbox{%
\begin{minipage}{0.99\columnwidth}
\textbf{Answer to RQ1:} The results indicate a strong correlation between attention allocation and hallucination sensitivity. Sections with lower attention, particularly S2R, exhibit higher susceptibility to hallucination. This insight directly informs the dataset design, ensuring that evaluation focuses on structurally challenging and under-attended components of bug reports.
\end{minipage}%
}
\end{center}

\subsection{Answering RQ2: Converting structured bug reports into high-quality unstructured reports reliably for training hallucination detection models}
\label{sub:RQ2}
\textbf{Evaluation Metric Justification.}  Evaluating structured-to-text generation is challenging because traditional lexical similarity metrics such as BLEU \cite{bleu} and ROUGE \cite{rouge} may not accurately reflect the similarity of the generated reports with the structured bug reports and penalize paraphrased outputs. In our setting, the requirements is not only lexical or semantic similarity but preserving of the information in the sections of structured bug reports. To address this limitation, we employ the PARENT metric and a variant of the PARENT metric $PARENT_{table}$, which evaluates factual alignment between structured input data and generated text. To validate the suitability of PARENT for this task, Pearson correlation coefficients were computed across all generated 10,304 reports. Table~\ref{tab:metric-correlation} presents the correlation between different evaluation metrics and PARENT scores. The results indicate moderate correlations between BLEU, ROUGE, and PARENT (0.61–0.67), suggesting that lexical overlap captures only part of the generation quality. In addition, BERTScore-F1 shows lower correlations with $PARENT_{bug}$ ($r=0.504$) and $PARENT_{table}$ ($r=0.471$), while correlating more strongly with ROUGE-1 ($r=0.740$) and ROUGE-L ($r=0.725$). This indicates that BERTScore captures contextual similarity between texts but does not fully substitute for source-field preservation. In contrast, source recall shows the strongest correlation with PARENT (0.73–0.79), indicating that PARENT more closely reflects the preservation of structured bug information during conversion. Therefore, we use PARENT as the primary metric for dataset curation.
\begin{table}[t]
\centering
\caption{Correlation of different evaluation metrics with $\mathrm{PARENT}$ and $\mathrm{PARENT}_{table}$ scores.}
\label{tab:metric-correlation}

\begin{tabular}{lcc}
\textbf{Metric} & \textbf{$\mathrm{PARENT}$ Corr.} & \textbf{$\mathrm{PARENT}_{table}$ Corr.} \\
\midrule
BLEU & 0.654 & 0.660 \\
ROUGE-1 & 0.675 & 0.643 \\
ROUGE-L & 0.640 & 0.615 \\
Cosine TF-IDF & 0.645 & 0.716 \\
BERTScore-F1 & 0.504 & 0.471 \\
Reference Recall & 0.251 & 0.222 \\
Source Recall & \textbf{0.787} & \textbf{0.735} \\
\end{tabular}
\vspace{-5mm}
\end{table}
We further inspected row-level cases where BERTScore-F1 was high but $PARENT_{table}$ was low. These cases show that generated reports can remain semantically similar to the source while omitting or weakening source-field information from S2R, AB, or EB. Therefore, BERTScore alone is insufficient for filtering structured-to-text conversions. The inspected examples are included in \footnote{\url{https://anonymous.4open.science/r/Anon_Bug_Report_Summarization-1477/unstructured_conversion/Data/final_conversion/metric_comparison/metric_comparison_bert_vs_parent_examples.md}}

\textbf{Generation Configuration.} After establishing PARENT as the primary evaluation metric, generation parameters were tuned to maximize the preservation of structured bug information in the generated reports. The $PARENT_{table}$ score was used as the optimization metric. Three prompt formulations were evaluated: a \textit{Baseline Prompt} adapted from prior work \cite{bug_report_llm_2025} which rewrites the report into a paragraph, a \textit{Natural Narrative Prompt} designed to produce fluent human-like descriptions, and a more constrained, and \textit{Strict Prompt} that emphasizes strict preservation of the original wording and structured information. We use these prompts to compare natural rewriting against stricter source-faithful conversion. Among these, the \textit{Strict Prompt} produced the more consistent $PARENT_{table}$ scores (mean $=0.564$, median $=0.575$) and was selected for dataset generation. Sampling temperature was varied between $0.0$ and $1.0$ to analyze the effect of generative variability on factual consistency, with lower temperatures yielding slightly higher $PARENT_{table}$ scores; a temperature of $0.2$ was therefore adopted. Finally, the generation length was dynamically scaled according to the source report length to ensure sufficient coverage of structured content while maintaining concise outputs. 

\begin{figure}[t]
\centering
\includegraphics[width=0.48\columnwidth]{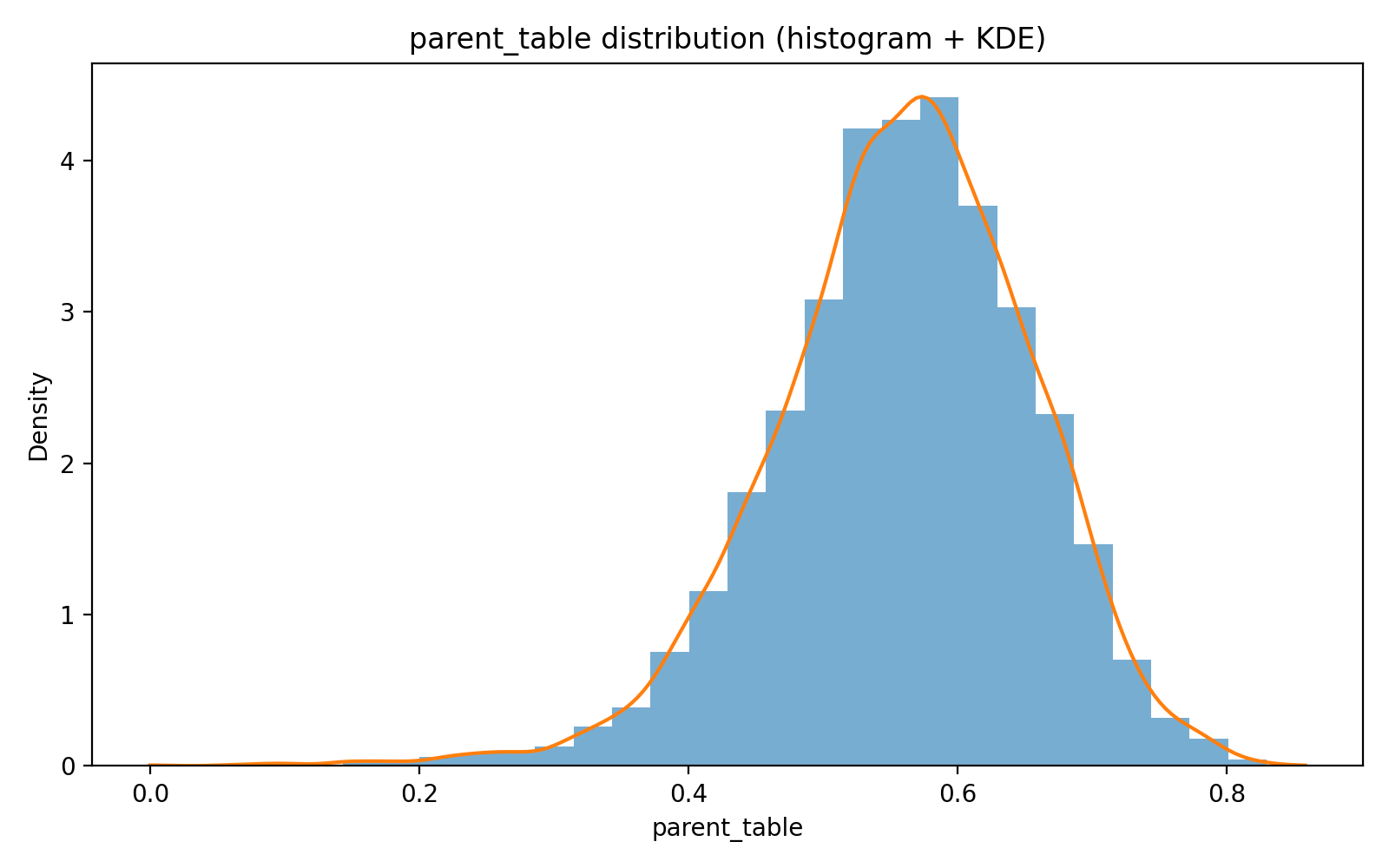}
\hfill
\includegraphics[width=0.48\columnwidth]{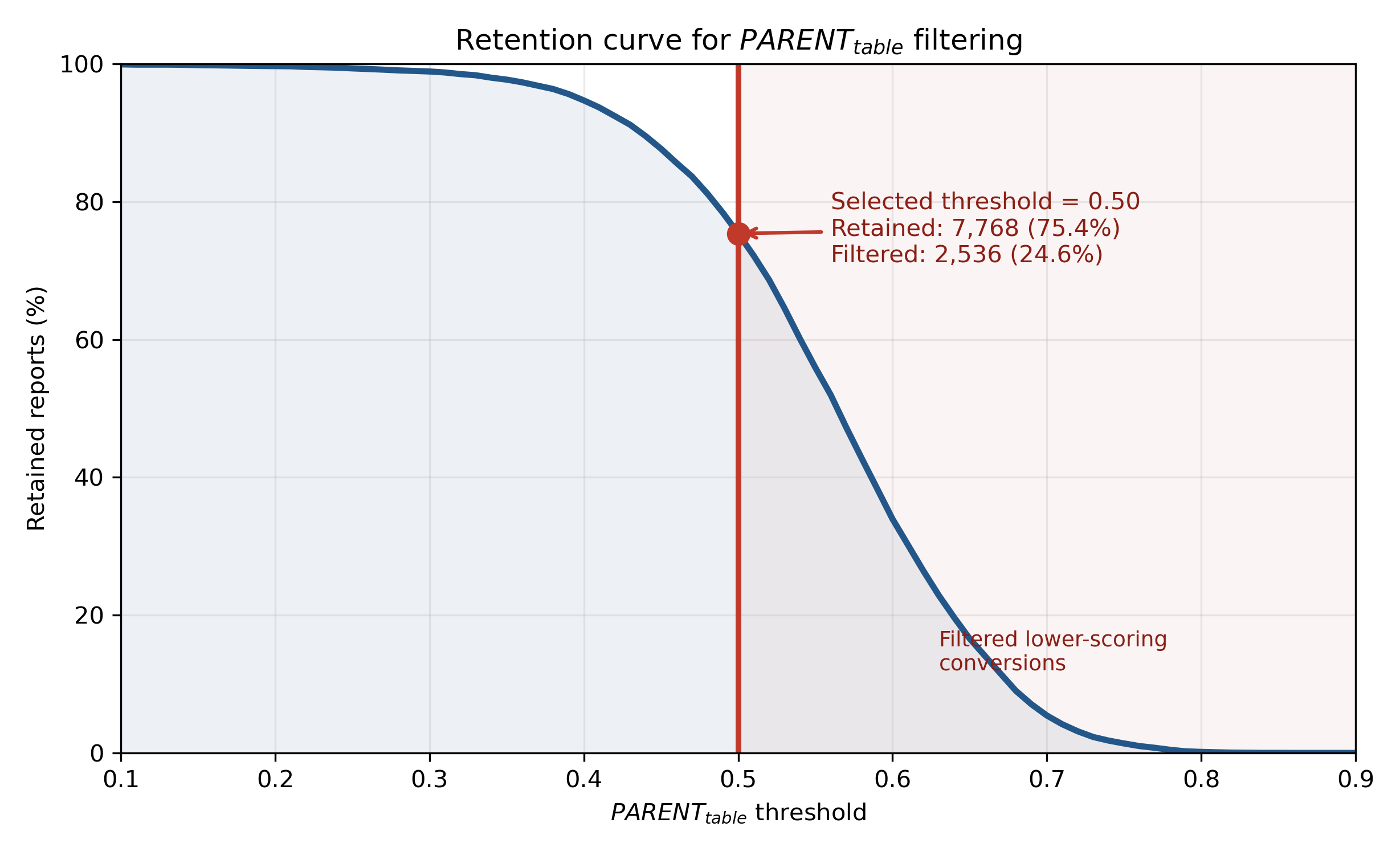}
\caption{Quality analysis of generated bug report descriptions. 
(a) Distribution of $\mathrm{PARENT}_{table}$ scores across generated reports, showing that most conversions preserve structured bug information. 
(b) Filtering strategy based on the score distribution, where low-scoring outliers are removed to ensure dataset reliability.}
\label{fig:dataset-quality}
\vspace{-8mm}
\end{figure}

\textbf{Conversion Quality and Dataset Filtering.}
After generating the narrative bug reports using the selected configuration (Strict prompt, 0.2 temperature, and dynamic generation length), the quality of the conversions was analyzed using the $PARENT_{table}$ metric. Figure~\ref{fig:dataset-quality}(a) shows the distribution of $PARENT_{table}$ scores across the generated dataset, indicating that around 75\% reports achieve moderate to high scores and therefore preserve the structured bug information effectively. However, a portion of the samples exhibit lower scores, corresponding to cases where important structured fields are partially omitted or insufficiently expressed in the generated text. \\
To ensure dataset reliability, low-quality conversions were removed using a filtering threshold of $PARENT_{table} \geq 0.50$. As shown in Figure~\ref{fig:dataset-quality}(b), this threshold retained 7,768 reports (75.39\%) and filtered out 2,536 lower-scoring reports (24.61\%). The selected threshold lies near the beginning of the steep retention-loss region, providing a more conservative filter than a relaxed threshold. This results in a filtered dataset that maintains strong semantic consistency with the original structured bug reports.
\begin{center}
\setlength{\fboxsep}{6pt}
\setlength{\fboxrule}{0.4pt}
\fbox{%
\begin{minipage}{0.99\columnwidth}
\textbf{Answer to RQ2:} The evaluation confirms that $\mathrm{PARENT}_{table}$ is a suitable metric for structured-to-text bug report conversion. Using $\mathrm{PARENT}_{table}$ to guide prompt selection, temperature tuning, and filtering, the generation pipeline produces unstructured reports that preserve the structured information from the original bug reports.
\end{minipage}%
}
\end{center}
\subsection{Answering RQ3: Multi-task learning model to detect hallucinations, identify affected sections, and classify hallucination types.}
\label{sub:RQ3}
To evaluate the effectiveness of the proposed section-aware hallucination detection approach, we fine-tuned the models using the constructed synthetic dataset. The dataset was divided into training, validation, and test sets using a 70\%, 10\%, 20\% split for fine-tuning, model fitting, and final evaluation, respectively. Experiments were conducted using four pretrained language models with different parameter scales: Mini-LLaMA-1B, Ministral-3B, Mistral-7B, and LLaMA-8B, which were selected to evaluate the impact of model scale on section-aware hallucination detection performance under practical computational constraints. The multitask model jointly predicts three tasks: (i) report-level hallucination detection, (ii) section-level hallucination localization, and (iii) hallucination type classification. The performance is evaluated using accuracy and F1 metrics with task-specific confusion matrices.

\begin{table*}[t]
\centering
\caption{Overall performance comparison of prompting and fine-tuned section-aware hallucination detection models.}
\label{tab:rq3-overall}
\begin{adjustbox}{width=\textwidth}
\begin{tabular}{lllccccc}
\toprule
Model & Approach & Prompt & Report Acc. & Report Macro-F1 & Section Micro-F1 & Section Macro-F1 & Type Macro-F1 \\
\midrule

\multirow{3}{*}{LLaMA-8B}
& Prompting & Zero-shot & 0.493 & 0.480 & 0.244 & 0.230 & 0.277 \\
& Prompting & Few-shot  & 0.550 & 0.550 & 0.398 & 0.407 & 0.295 \\
& Prompting & CoT       & 0.422 & 0.314 & 0.327 & 0.310 & 0.182 \\
\cmidrule(lr){1-8}

\multirow{3}{*}{Mistral-7B}
& Prompting & Zero-shot & 0.578 & 0.427 & 0.073 & 0.072 & 0.218 \\
& Prompting & Few-shot  & 0.418 & 0.308 & 0.281 & 0.198 & 0.193 \\
& Prompting & CoT       & 0.461 & 0.457 & 0.231 & 0.168 & 0.205 \\
\cmidrule(lr){1-8}

\multirow{3}{*}{Ministral-3B}
& Prompting & Zero-shot & 0.683 & 0.679 & 0.212 & 0.212 & 0.484 \\
& Prompting & Few-shot  & 0.662 & 0.661 & 0.299 & 0.270 & 0.454 \\
& Prompting & CoT       & 0.640 & 0.639 & 0.234 & 0.224 & 0.477 \\
\cmidrule(lr){1-8}

Mini-LLaMA   & Fine-tuned & N/A & 0.832 & 0.816 & 0.716 & 0.723 & 0.662 \\
Ministral-3B & Fine-tuned & N/A & \textbf{0.898} & \textbf{0.891} & \textbf{0.846} & \textbf{0.829} & \textbf{0.836} \\
Mistral-7B   & Fine-tuned & N/A & 0.889 & 0.881 & 0.832 & 0.809 & 0.817 \\
LLaMA-8B     & Fine-tuned & N/A & 0.892 & 0.885 & 0.839 & 0.820 & 0.802 \\

\bottomrule
\end{tabular}
\end{adjustbox}
\end{table*}
\noindent\textbf{Model Performance.} Table~\ref{tab:rq3-overall} summarizes the performance of the proposed section-aware hallucination detection approach across all fine-tuned models and prompting baselines. The results demonstrate that compared to zero-shot, few-shot, and CoT prompting, all fine-tuned models achieve strong performance (F1-score > 0.8) across the three tasks, including report-level hallucination detection, section-level hallucination localization, and hallucination type classification. \\
For report-level hallucination detection, Ministral-3B achieved the best performance, with an accuracy of 0.898 and a macro-F1 score of 0.891. Mistral-7B and LLaMA-8B also achieved comparable performance, with macro-F1 scores of 0.881 and 0.885, respectively. Mini-LLaMA-1.1B achieved lower performance, reflecting the limited representational capacity of smaller models. For section-level hallucination localization, Ministral-3B again obtained the highest performance, with a micro-F1 score of 0.846 and a macro-F1 score of 0.829. Mistral-7B and LLaMA-8B followed closely, while Mini-LLaMA-1.1B showed a larger performance gap. The improvement over Mini-LLaMA suggests that richer contextual representations help the model better capture structural inconsistencies across sections.\\
For hallucination type classification, Ministral-3B achieved the best macro-F1 score of 0.836, outperforming Mistral-7B and LLaMA-8B. This indicates that increasing model size alone does not necessarily lead to better hallucination-type classification. 
Overall, the results indicate that the proposed section-aware learning approach can effectively detect hallucinated bug report summaries while simultaneously identifying the affected sections and the type of hallucination. Although the 3B, 7B, and 8B models achieve broadly comparable performance, Ministral-3B provides the strongest overall results while requiring fewer computational resources than the larger models. \\
\textbf{Section-Level Hallucination Localization.}
\begin{table}[t]
\centering
\caption{Detailed performance of the best-performing model, Ministral-3B: (a) section localization and (b) hallucination type classification.}
\label{tab:rq3-detailed}
\centering
\textbf{(a) Section localization}\\[1mm]
\begin{tabular*}{\columnwidth}{@{\extracolsep{\fill}}lccc}
\toprule
Section & Prec. & Rec. & F1 \\
\midrule
S2R & 0.935 & \textbf{0.851} & \textbf{0.891} \\
AB  & 0.989 & 0.656 & 0.789 \\
EB  & \textbf{1.000} & 0.675 & 0.806 \\
\end{tabular*}
\hfill
\centering
\textbf{(b) Type classification}\\[1mm]
\begin{tabular*}{\columnwidth}{@{\extracolsep{\fill}}lccc}
\toprule
Type & Prec. & Rec. & F1 \\
\midrule
Add & \textbf{1.000} & \textbf{0.954} & \textbf{0.976} \\
Remove & 0.831 & 0.383 & 0.525 \\
Reorder & 0.963 & 0.888 & 0.924 \\
\end{tabular*}
\end{table}
To further analyze the model in localizing hallucinations, we evaluated section-level detection performance for the best-performing model, Ministral-3B. Table~\ref{tab:rq3-detailed}(a) reports the precision, recall, and F1-scores for detecting hallucinations in each bug report section. The model achieved the highest performance for the S2R section with an F1 of 0.891. Performance was slightly lower for the AB and EB sections, which obtained F1-scores of 0.789 and 0.806, respectively. Across all sections, the model achieved high precision, especially for AB and EB, but recall was lower for these sections. This suggests that the model is conservative when assigning hallucinations to AB and EB while it predicts these sections, it is usually correct, but it misses some hallucinations in those sections. This difference may be attributed to the more structured nature of reproduction steps, which provide clearer contextual insights for identifying hallucinated content.
\\ \textbf{Hallucination Type Classification.} To evaluate whether the model can distinguish different hallucination patterns, we analyze the hallucination type classification performance of the best-performing model, Ministral-3B. As shown in Table~\ref{tab:rq3-detailed}(b), the model achieves strong performance for addition and reordering hallucinations, achieving high precision and recall.
In contrast, removal-type hallucinations are more difficult to detect, resulting in lower recall. This is expected, as removal errors correspond to missing information rather than explicit textual additions, making them harder to identify. Despite this challenge, Ministral-3B improves removal detection compared with the larger models and achieves strong overall performance across hallucination types. \\
\textbf{Qualitative Error Analysis.}
To better understand the limitations of the proposed detector, we analyzed all incorrect predictions (163 cases) produced by the best-performing model, Ministral-3B, from the 1,354 synthetic test cases. The error set was dominated by removal hallucinations, particularly in declarative sections with AB-removal and EB-removal accounted for 42 and 44 incorrect cases, respectively. We manually coded each incorrect prediction using explanation labels that capture why the prediction failed, including \textit{missing or implicit information}, \textit{procedural confusion}, \textit{section-boundary ambiguity}, \textit{technical context loss}, \textit{paraphrase over-triggering}, \textit{long/multi-issue complexity}, and \textit{synthetic label ambiguity}. To support reliability, 30 cases were independently annotated by three authors, while the remaining 133 cases were distributed. The overlap subset was stratified by ground-truth section and hallucination type. The Fleiss' kappa was 0.200, and all conflicting cases were resolved through discussion.
\begin{figure}[t]
\centering
\includegraphics[width=0.9\columnwidth]{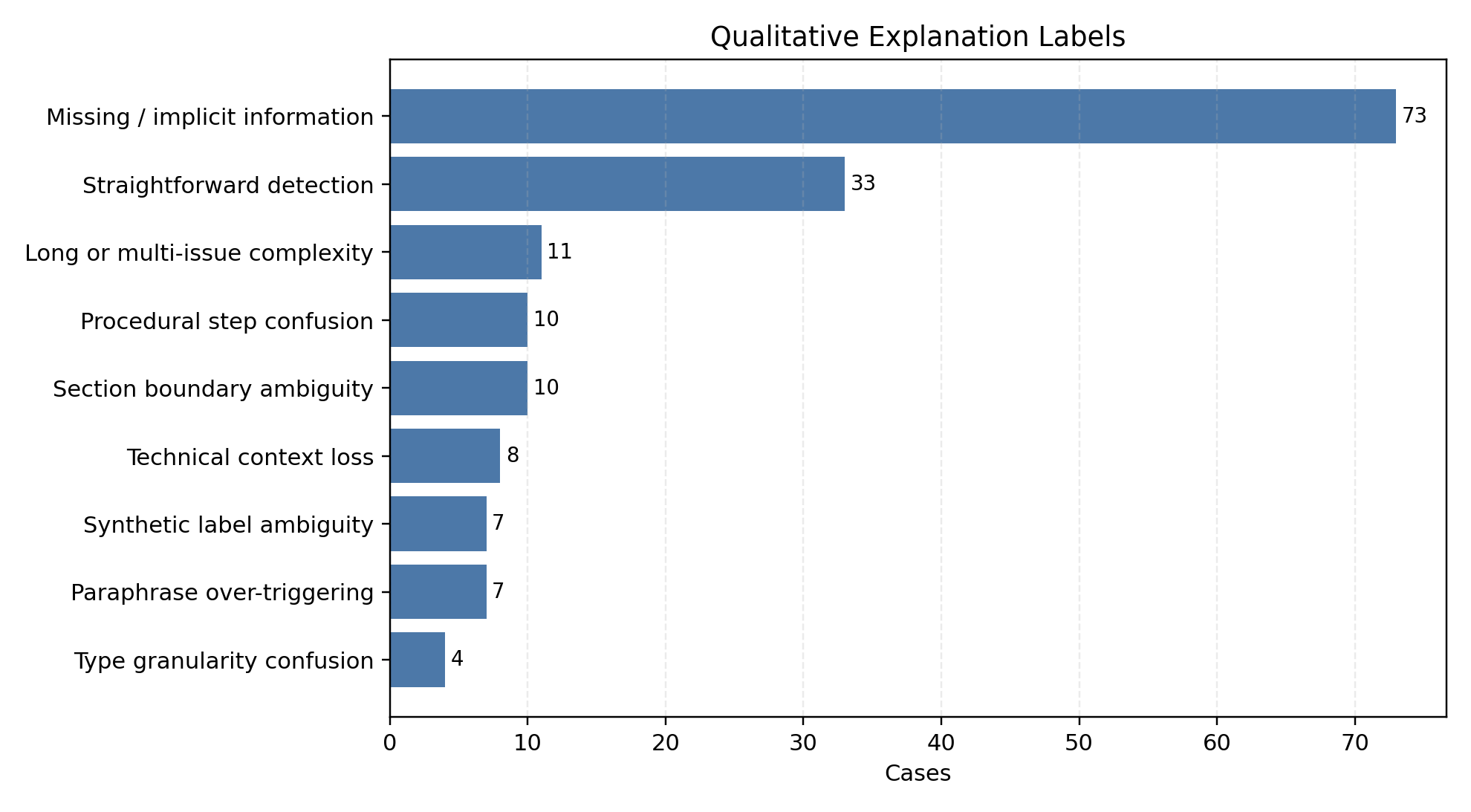}
\caption{Frequencies of manual explanation labels from the qualitative error analysis.}
\label{fig:manual-explanation-labels}
\vspace{-7mm}
\end{figure}
Figure \ref{fig:manual-explanation-labels} summarizes the explanation labels distribution. The most frequent explanation was \textit{missing or implicit information hard to notice}, accounting for 73 of 163 cases (44.8\%), followed by \textit{straightforward detection failure} with 20.2\%, while the remaining categories each accounted for less than 7\% of the errors. The dominant failure mode was \textit{missed omitted or implicit information}, appearing in 60.7\% of removal cases, 54.6\% of report-level false negatives, and 50.5\% of high-confidence wrong predictions. The pattern was more common in EB and AB than in S2R, indicating that removal hallucinations are harder because the model must detect absent source information rather than visible unsupported text. False positives were more heterogeneous, often reflecting \textit{paraphrase over-triggering}, \textit{straightforward detection failures}, or \textit{borderline section ambiguity}. 
Overall, the qualitative analysis shows that the detector's main limitation is not general hallucination recognition, but detecting omitted or implicitly preserved information, especially in removal hallucinations.
\begin{center}
\setlength{\fboxsep}{6pt}
\setlength{\fboxrule}{0.4pt}
\fbox{%
\begin{minipage}{0.9\columnwidth}
\textbf{Answer to RQ3:}
The proposed section-aware approach effectively detects hallucinated bug report summaries while identifying the affected sections and hallucination types. Ministral-3B achieved the best overall performance, showing that strong  detection can be obtained with a resource-efficient mid-scale model. Error analysis shows that removal hallucinations, especially in AB and EB sections, remain the most challenging case. The results show that multitask learning enables more detailed and interpretable hallucination detection than traditional binary approaches.
\end{minipage}%
}
\end{center}

\section{Discussion and Future work}
\label{sec:Discussion}
The findings of this study highlight the importance of incorporating document structure when analyzing hallucinations in LLM-generated bug report summaries. The exploratory analysis demonstrated that hallucination patterns vary across semantic sections, suggesting that different contexts influence the summary generation. The multitask learning approach shows that jointly modeling report-level hallucination detection, section identification, and hallucination type classification can provide richer signals for detecting factual inconsistencies. In our experiments, the best-performing model achieved a report-level accuracy of 0.892, while obtaining a section-level micro-F1 of 0.839, demonstrating the effectiveness of the proposed section-aware detection approach.

Despite these promising results, several directions remain for future research. First, the current evaluation focuses on a limited set of model scales due to computational constraints. Future work could investigate performance trends across medium and larger models (e.g., Llama-17B) to better understand the relationship between model scale and hallucination detection capabilities. Second, while this work relies on fine-tuned classifiers, it would be valuable to compare the proposed approach with zero-shot and few-shot prompting strategies using LLMs. Third, detecting missing or incomplete information remains challenging and requires further investigation, particularly for removal-type hallucinations. Finally, although synthetic hallucination injection enables controlled experimentation, validating the approach on manually annotated real-world datasets would provide stronger evidence of its applicability in practical SE workflows.

\section{Threats to Validity}
\label{sec:Threats}
\textbf{Construct Validity}
\label{subsec:construct_validity}
A primary threat involves our hallucination taxonomy (addition, removal, and reordering), which was established via an exploratory analysis of 80 reports. While this covers prominent error modes, it may not encompass all subtle or domain-specific factual inconsistencies. Furthermore, due to the lack of pre-existing human-annotated bug report hallucination corpora, we rely on an LLM-driven synthetic injection pipeline to train and evaluate our models. These synthetic perturbations serve as an operational proxy and might not perfectly mirror the semantic nuances or distribution of organic hallucinations produced in real-world software workflows. Finally, our report filtering relies on the modified $\text{PARENT}_{\text{table}}$ metric, which emphasizes token overlap and might mischaracterize highly creative yet faithful paraphrases. To mitigate these threats, we validated our exploratory baseline with double-blind annotations achieving high agreement (Cohen's $\kappa = 0.85$) and implemented automated validation rules during synthesis to ensure the structural and behavioral consistency of injected errors.\\
\textbf{Internal Validity}
\label{subsec:internal_validity}
A main threat relates to using self-attention scores as an operational proxy for model interpretability; in deep learning architectures, attention weights do not always map linearly to underlying model reasoning or provide an exhaustive explanation of prediction behavior. Additionally, our text preprocessing and artifact engineering introduce configuration bias; specifically, the insertion of structural delimiters used to isolate fields (such as \texttt{[S2R]}, \texttt{[AB]}, and \texttt{[EB]}) fundamentally alters sequence lengths and input token distributions, directly influencing how attention weights are distributed by the transformer layers. Finally, the choice of specific models (e.g., Mistral-7B for conversion and Ministral-3B for multi-task detection) introduces model-specific configuration bias. To control for these threats, we enforced a strict, uniform data-cleaning and tokenization pipeline across all configurations. Furthermore, rather than relying on a single architecture, we evaluated our section-level attention analysis across five distinct transformer models (BERT, RoBERTa, SciBERT, DeBERTa, and DistilBERT) to isolate reproducible behavioral trends rather than design-specific artifacts.\\
\textbf{External Validity}
\label{subsec:external_validity}
In our study, the foremost limitation lies in the structural and domain diversity of our subject systems, as our evaluation is bounded by the open-source characteristics of the BugsRepo dataset. Consequently, our findings and fine-tuned models might not immediately generalize to closed-source enterprise tracking ecosystems (e.g., Jira, ServiceNow) or highly customized, unstructured issue-tracking workflows that lack clear section boundaries. Moreover, the specific syntax of our injection vectors and the mathematical constraints of the evaluation metrics heavily shape performance outcomes, meaning model sensitivity could vary under alternative prompting styles or compression ratios. We addressed the immediate risk of overfitting and ensured statistical stability by executing our multi-task detection experiments on a sufficiently large validation partition that provides stable metric responses. To reinforce external generalizability, our future work involves expanding this benchmark to encompass heterogeneous, multi-domain software engineering datasets and validating our section-aware models in production-grade software environments.

\section{Conclusion}
\label{sec:Conclusion}
This empirical study investigated hallucinations in LLM-generated bug report summaries through a section-aware perspective using open source software repository dataset. We first conducted an exploratory analysis showing that hallucination patterns vary across different semantic sections of bug reports. Based on these observations, we constructed a synthetic benchmark by identifying less-attended bug report sections, injecting controlled hallucinations, converting structured bug reports into unstructured narratives, and filtering low-quality generations using a PARENT-based scoring strategy. We then proposed a section-aware hallucination detection approach capable of simultaneously identifying hallucinated summaries, affected report sections, and hallucination types. This empirical study results demonstrated that our proposed approach effectively supports fine-grained hallucination detection while providing concrete interpretable insights. Overall, our findings suggest that incorporating document structure into hallucination detection can improve the reliability and transparency of LLM-assisted bug report summarization systems in SE workflows. Thus, our study provides empirical evidence of hallucination risks in LLM-assisted bug-report summarization, with a detection framework for improving the reliability of software engineering artifacts.

\section{Data Availability}
\label{sec:DataAvailability}
To support the reproducibility and verifiability of our study, the datasets used and generated in this work are publicly available across two primary repositories:
\begin{itemize}
    \item \textbf{Source Bug Reports:} The underlying structured bug reports used for our exploratory analysis and synthetic generation are sourced from the \textit{BugsRepo} dataset, which is permanently archived and publicly accessible on Zenodo at \url{https://zenodo.org/records/15004067}.
    \item \textbf{Replication Package and Processed Data:} The complete replication package, including the controlled synthetic hallucination injection scripts, the finalized section-aware benchmark dataset, the trained model checkpoints, and evaluation scripts, is hosted anonymously to preserve the double-blind review process. It can be accessed at \url{https://anonymous.4open.science/r/Anon_Bug_Report_Summarization-1477/data}. 
\end{itemize}

\bibliography{references}
\end{document}